\documentclass[]{spie}

\usepackage{amsmath,amssymb,amsfonts}
\usepackage{graphicx}
\usepackage{subcaption}
\usepackage{float}
\usepackage{authblk}

\title{Aliased noise characterization and mitigation in BICEP Array 150, 220 and 270 GHz time-division multiplexed detectors}

\author[a]{S.~Fatigoni}
\author[b]{P.~A.~R.~Ade}
\author[c,d]{Z.~Ahmed}
\author[e]{M.~Amiri}
\author[f]{D.~Barkats}
\author[a]{R.~Basu~Thakur}
\author[g]{D.~Beck}
\author[h]{C.~A.~Bischoff}
\author[a,i]{J.~J.~Bock}
\author[j]{V.~Buza}
\author[g,c]{B.~Cantrall}
\author[a]{J.~R.~Cheshire~IV}
\author[k]{J.~Connors}
\author[l]{J.~Cornelison}
\author[m]{M.~Crumrine}
\author[a]{A.~J.~Cukierman}
\author[k]{E.~Denison}
\author[n]{L.~Duband}
\author[f]{M.~A.~Echter}
\author[o]{M.~Eiben}
\author[f,p]{B.~D.~Elwood}
\author[q]{J.~P.~Filippini}
\author[g]{A.~Fortes}
\author[a]{M.~Gao}
\author[h]{C.~Giannakopoulos}
\author[g]{N.~Goeckner-Wald}
\author[g]{D.~C.~Goldfinger}
\author[r,s]{S.~Gratton}
\author[g]{J.~A.~Grayson}
\author[a]{A.~Greathouse}
\author[f]{P.~K.~Grimes}
\author[e]{M.~Halpern}
\author[c,d]{S.~Henderson}
\author[m]{T.~D.~Hoang}
\author[k]{J.~Hubmayr}
\author[a]{H.~Hui}
\author[g]{K.~D.~Irwin}
\author[t]{M.~Izquierdo~Poza}
\author[a]{J.~H.~Kang}
\author[t]{K.~S.~Karkare}
\author[a]{S.~Kefeli}
\author[f,p]{J.~M.~Kovac}
\author[g]{C.~Kuo}
\author[m,u]{K.~Lasko}
\author[a]{K.~Lau}
\author[h]{M.~Lautzenhiser}
\author[g]{T.~Liu}
\author[j,v]{S.~C.~Mackey}
\author[m]{N.~Maher}
\author[i]{K.~G.~Megerian}
\author[a]{L.~Minutolo}
\author[a]{L.~Moncelsi}
\author[g]{Y.~Nakato}
\author[a,i]{H.~T.~Nguyen}
\author[a,i]{R.~O'Brient}
\author[f]{S.~N.~Paine}
\author[a]{A.~Patel}
\author[f]{M.~A.~Petroff}
\author[f,p]{A.~R.~Polish}
\author[n]{T.~Prouve}
\author[m]{C.~Pryke}
\author[k]{C.~D.~Reintsema}
\author[a]{T.~Romand}
\author[g]{M.~Salatino}
\author[a]{A.~Schillaci}
\author[f]{B.~Schmitt}
\author[m,u]{B.~Singari}
\author[a,i]{A.~Soliman}
\author[f]{T.~St.~Germaine}
\author[a]{A.~Steiger}
\author[a]{B.~Steinbach}
\author[b]{R.~Sudiwala}
\author[g,c]{K.~L.~Thompson}
\author[b]{C.~Tucker}
\author[i]{A.~D.~Turner}
\author[w]{C.~Verg\`{e}s}
\author[j,v]{A.~G.~Vieregg}
\author[a]{A.~Wandui}
\author[i]{A.~C.~Weber}
\author[m]{J.~Willmert}
\author[c,d]{W.~L.~K.~Wu}
\author[g]{H.~Yang}
\author[j,l]{C.~Yu}
\author[f]{L.~Zeng}
\author[c]{C.~Zhang}
\author[a]{S.~Zhang}

\affil[a]{Department of Physics, California Institute of Technology, Pasadena, CA 91125, USA}
\affil[b]{School of Physics and Astronomy, Cardiff University, Cardiff, CF24 3AA, United Kingdom}
\affil[c]{Kavli Institute for Particle Astrophysics and Cosmology, Stanford University, Stanford, CA 94305, USA}
\affil[d]{SLAC National Accelerator Laboratory, Menlo Park, CA 94025, USA}
\affil[e]{Department of Physics and Astronomy, University of British Columbia, Vancouver, British Columbia, V6T 1Z1, Canada}
\affil[f]{Center for Astrophysics, Harvard \& Smithsonian, Cambridge, MA 02138, USA}
\affil[g]{Department of Physics, Stanford University, Stanford, CA 94305, USA}
\affil[h]{Department of Physics, University of Cincinnati, Cincinnati, OH 45221, USA}
\affil[i]{Jet Propulsion Laboratory, California Institute of Technology, Pasadena, CA 91109, USA}
\affil[j]{Kavli Institute for Cosmological Physics, University of Chicago, Chicago, IL 60637, USA}
\affil[k]{National Institute of Standards and Technology, Boulder, CO 80305, USA}
\affil[l]{Argonne National Laboratory, High Energy Physics Division, Lemont, IL 60439, USA}
\affil[m]{School of Physics and Astronomy, University of Minnesota, Minneapolis, MN 55455, USA}
\affil[n]{Service des Basses Temperatures, Commissariat a l’Energie Atomique, 38054 Grenoble, France}
\affil[o]{Faculty of Physical Sciences, University of Iceland, 102 Reykjavík, Iceland}
\affil[p]{Department of Physics, Harvard University, Cambridge, MA 02138, USA}
\affil[q]{Department of Physics, Grainger College of Engineering, University of Illinois Urbana-Champaign, Urbana, IL 61801, USA}
\affil[r]{Centre for Theoretical Cosmology, University of Cambridge, Cambridge, CB3 0WA, United Kingdom}
\affil[s]{Kavli Institute for Cosmology Cambridge, University of Cambridge, Cambridge, CB3 0HA, United Kingdom}
\affil[t]{Department of Physics, Boston University, Boston, MA 02215, USA}
\affil[u]{Minnesota Institute for Astrophysics, University of Minnesota, Minneapolis, MN 55455, USA}
\affil[v]{Department of Physics, Astronomy \& Astrophysics, Enrico Fermi Institute, University of Chicago, Chicago, IL 60637, USA}
\affil[w]{Physics Division, Lawrence Berkeley National Laboratory, Berkeley, CA 94720, USA}

\authorinfo{Further author information: \\
S.F.: sofiaf@caltech.edu}

\begin{document}

\maketitle

\begin{abstract}

Early observations with the BICEP Array 150 GHz (BA2-150) and 220/270 GHz (BA3-220/270) receivers revealed detector noise equivalent temperatures (NETs) higher than expected, together with substantial detector-to-detector and module-to-module scatter. Noise measurements acquired with multiplexing off and high frequency sampling demonstrate that this excess originates from elevated high-frequency detector noise that aliases into the science band during time-division multiplexing.

We show that the excess high-frequency noise is correlated with anomalously large logarithmic TES transition slopes, $\alpha$, resulting in elevated electrothermal loop gain and operation near the detector stability boundary. Measurements of $\alpha$ indicate values substantially larger than expected, consistent with the sharper superconducting transitions introduced by the inverted TES fabrication process adopted for BA2-150 and BA3-220/270 detectors. 

Operational mitigation strategies were investigated through both increased multiplexing rates and elevated focal-plane operating temperatures. Faster multiplexing reduces aliasing by shifting the multiplexing Nyquist frequency beyond the excess noise roll-off, while elevated bath temperatures reduce TES electrical power and loop gain, improving detector stability and reducing NET by approximately 10\%.

These results demonstrate the importance of balancing TES responsivity, electrothermal stability, and multiplexed readout performance in next-generation CMB polarimeters.


\end{abstract}

\section{Introduction}

The theory of Cosmic Inflation posits that the universe underwent a period of rapid expansion during the first fractions of a second following the Big Bang. One of the primary observational signatures of inflation is a primordial B-mode polarization pattern in the Cosmic Microwave Background (CMB). Detecting this signal would provide strong evidence for Inflation and significantly constrain the parameter space of inflationary models \cite{BK2021}. Achieving the sensitivity required for such measurements demands large-format arrays of low-noise superconducting detectors operating across multiple observing frequencies in order to separate cosmological signal from polarized Galactic foreground emission.

The BICEP/Keck program is a series of small-aperture microwave telescopes located at the South Pole designed to measure large-angular-scale CMB polarization with high sensitivity\cite{Hui2018}. The latest instrument in the series, BICEP Array (BA), consists of multiple receivers operating between 30 and 270 GHz, each employing antenna-coupled transition-edge sensor (TES) bolometers \cite{IrwinHilton2005} read out through time-division multiplexed (TDM) SQUID electronics\cite{Fatigoni2024}.

This work focuses on the BA2-150 and BA3-220/270 receivers, which are the two highest frequency receivers. The BA2-150 and BA3-220/270 detectors employ an inverted TES fabrication process introduced to reduce magnetic pickup and improve fabrication robustness relative to earlier BICEP receivers\cite{Weber2024}. During the first observing seasons of these receivers, measured detector noise equivalent temperatures (NETs) were found to exceed expectations, with substantial scatter observed both between detector modules and between individual detectors within a module.

Detailed investigations suggested that the excess noise contribution originated from detector noise rather than photon noise alone. In particular, high-sampling-rate ($\sim$400~kHz) noise measurements revealed elevated high-frequency detector noise components that alias into the science band during multiplexed readout.

In this work, we investigate the physical origin of the excess aliased noise and its connection to TES loop-gain. We present measurements of TES transition sensitivity, studies of aliased noise as a function of multiplexing parameters and detector operating point, operational mitigation strategies involving optimized multiplexing rates and elevated focal-plane temperatures, and direct SQUID settling-time measurements constraining multiplexing optimization. Finally, we discuss future detector-design efforts aimed at reducing excess loop gain through TES transition engineering.

\section{TES Loop Gain, Stability, and Excess Noise}

TES bolometers are operated within the superconducting transition under voltage bias, where negative electrothermal feedback (ETF) stabilizes the detector response. A small increase in absorbed power raises the TES temperature and resistance, reducing the Joule heating and driving the detector back toward its operating point. This feedback linearizes the detector response, suppresses temperature fluctuations, and shortens the effective thermal response time.

\begin{figure}[H]
\centering

\begin{subfigure}[t]{0.42\textwidth}
    \centering
    \includegraphics[width=0.8\linewidth]{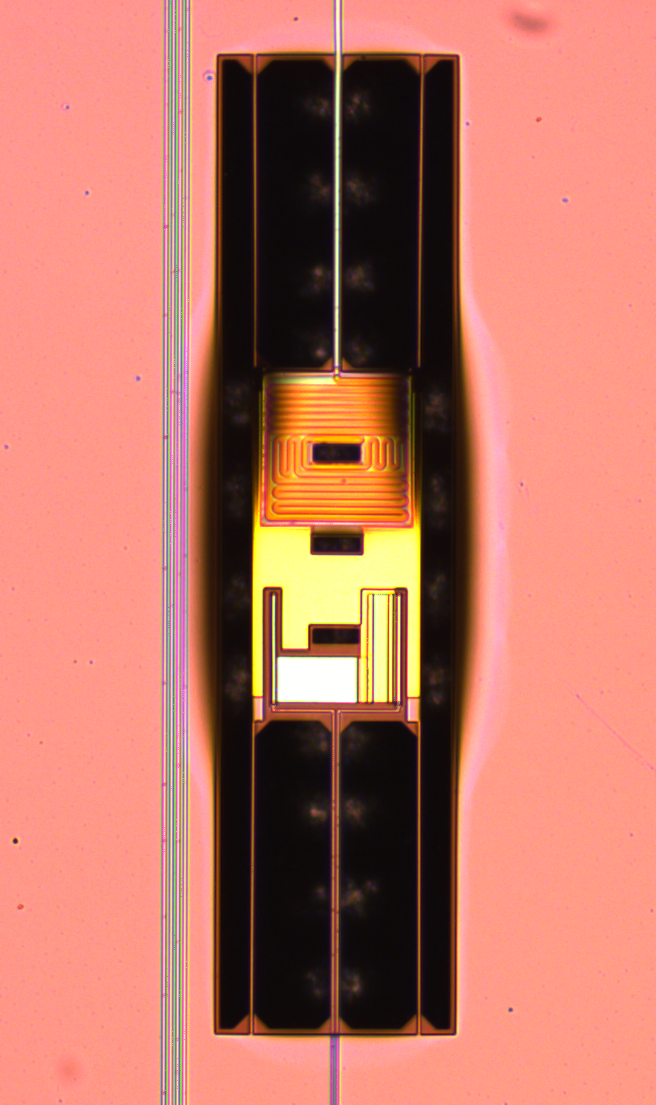}
    \caption{}
    \label{fig:tes_photo}
\end{subfigure}
\hfill
\begin{subfigure}[t]{0.52\textwidth}
    \centering
    \includegraphics[width=\linewidth]{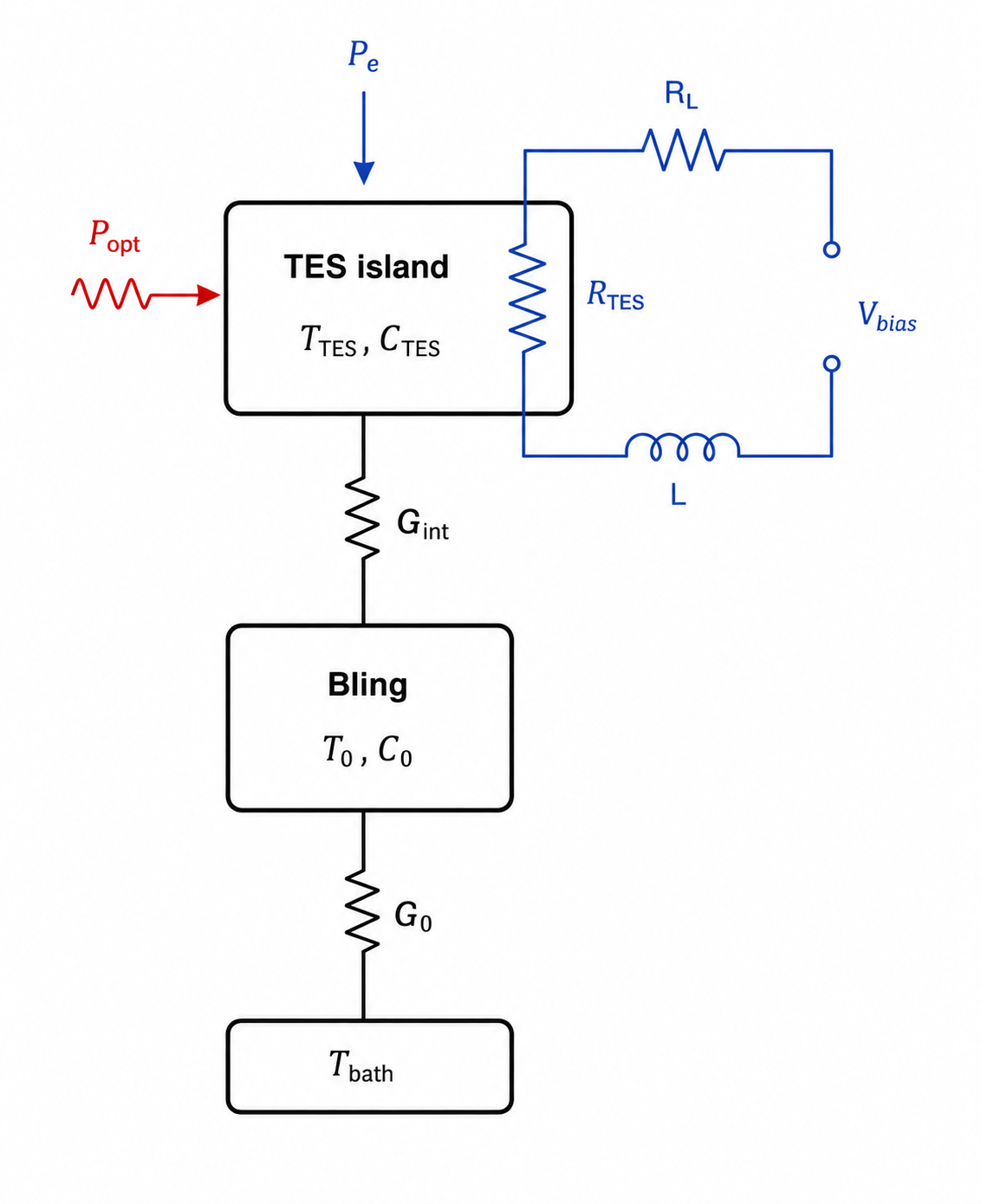}
    \caption{}
    \label{fig:tes_circuit}
\end{subfigure}
\vspace{0.5cm}
\caption{(a) Optical microscope image of the island of a BICEP Array 270 GHz TES detector. (b) Two-body thermal model of a TES bolometer, showing the thermal links between the TES, the auxiliary heat capacity (bling), and the thermal bath.}
\label{fig:tes_overview}

\end{figure}

The strength of the electrothermal feedback is quantified by the loop gain,

\begin{equation}
\mathcal{L}=\frac{\alpha P_e}{G_0T_c},
\end{equation}
\newline
where $P_e$ is the TES electrical power, $G_0$ is the thermal conductance to the bath (Fig.~\ref{fig:tes_overview}), $T_c$ is the superconducting transition temperature, and

\begin{equation}
\alpha=\frac{d\ln R}{d\ln T}
\end{equation}
\newline
is the logarithmic temperature sensitivity of the transition. A larger $\alpha$ increases detector responsivity and strengthens the electrothermal feedback, but also increases the loop gain.

The detector dynamics are governed by both the thermal and electrical response of the bias circuit. The natural thermal time constant is

\begin{equation}
\tau=\frac{C}{G},
\end{equation}
\newline
where $C$ is the TES heat capacity, and G is the differential thermal conductance to the bath $G=dP_{bath}/{dT}$.
The electrical time constant is given by

\begin{equation}
\tau_{\rm el}=\frac{L}{R_L+R_0(1+\beta_I)},
\end{equation}
\newline
where $L$ is the bias-circuit inductance, $R_L$ is the load resistance, $R_0$ is the TES operating resistance, and $\beta_I$ is the logarithmic current sensitivity of the TES resistance

\begin{equation}
\beta_{I}=\frac{d\ln R}{d\ln I}
\end{equation}
\newline
Electrothermal feedback reduces the effective thermal response time,

\begin{equation}
\tau_{\rm eff}\simeq\frac{\tau}{1+\mathcal{L}},
\end{equation}
\newline
allowing TES detectors to respond more rapidly. However, increasing the loop gain also reduces the separation between the thermal and electrical time constants. When this separation becomes too small, the detector becomes underdamped and approaches its electrothermal stability limit.

For a two-body thermal model (Fig.\ref{fig:tes_overview}), the stability criterion is approximately \cite{Sonka2018}

\begin{equation}
\mathcal{L}<\gamma+1,
\end{equation}
\newline
where

\begin{equation}
\gamma=\frac{G_{\rm int}}{G_0},
\end{equation}

and $G_{\rm int}$ is the thermal conductance between the TES and the auxiliary heat capacity ("bling"), while $G_0$ is the thermal conductance to the bath (see Fig.\ref{fig:tes_overview}). As $\mathcal{L}$ approaches this limit, the detector response becomes increasingly underdamped and electrothermal oscillations can develop. 

The fundamental noise sources in an ideal TES include thermal fluctuation noise across the weak thermal link, TES Johnson noise, load-resistor Johnson noise, and SQUID amplifier noise \cite{IrwinHilton2005}. In practice, however, TES devices often exhibit excess electrical noise beyond these ideal contributions \cite{Sonka2018}. Previous studies have shown that this excess noise is correlated with the sharpness of the superconducting transition and can be reduced through TES transition engineering, including the introduction of normal-metal structures that broaden the transition \cite{George2014}.

These observations motivate the interpretation that the unusually large $\alpha$ values measured in the BA2-150 and BA3-220/270 detectors produce elevated loop gain, driving many detectors close to the electrothermal stability boundary. Under these conditions, excess high-frequency detector noise is generated. During time-division multiplexing, this out-of-band noise aliases into the science band, increasing the measured detector NET. The proposed causal chain is therefore

\[
\mathrm{high}\ \alpha
\rightarrow
\mathrm{high}\ \mathcal{L}
\rightarrow
\mathrm{instability}
\rightarrow
\mathrm{excess\ high\text{-}frequency\ noise}
\rightarrow
\mathrm{aliased\ in\text{-}band\ noise}.
\]

This physical framework motivates the mitigation strategies explored in this work. Increasing the multiplexing rate raises the multiplexing Nyquist frequency, reducing the amount of high-frequency detector noise folded into the science band. Alternatively, reducing the TES electrical power lowers the loop gain directly. For the existing detectors this is achieved by operating at elevated bath temperatures, while future detector designs aim to reduce $\alpha$ through transition engineering, lowering the loop gain while preserving the fabrication and magnetic robustness advantages of the inverted TES architecture.

\section{Aliased High-Frequency Noise}

\begin{figure}[H]
\centering

\begin{subfigure}[t]{0.49\textwidth}
    \centering
    \includegraphics[width=\linewidth]{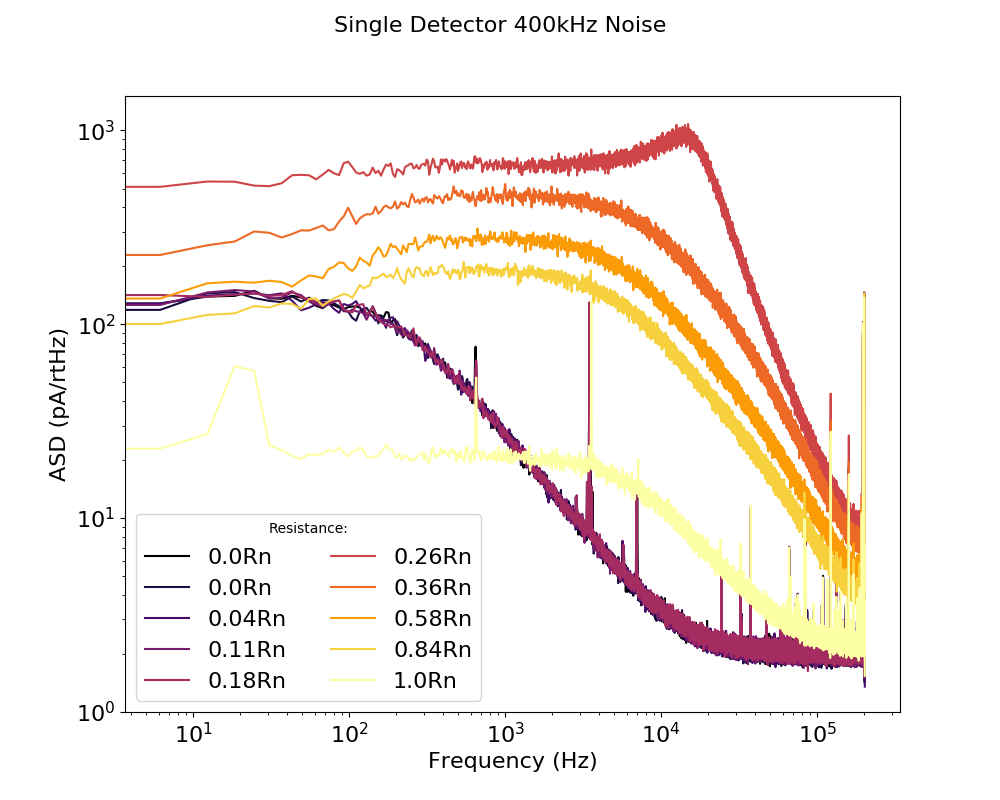}
    \caption{}
    \label{fig:noise_psd}
\end{subfigure}
\hfill
\begin{subfigure}[t]{0.49\textwidth}
    \centering
    \includegraphics[width=\linewidth]{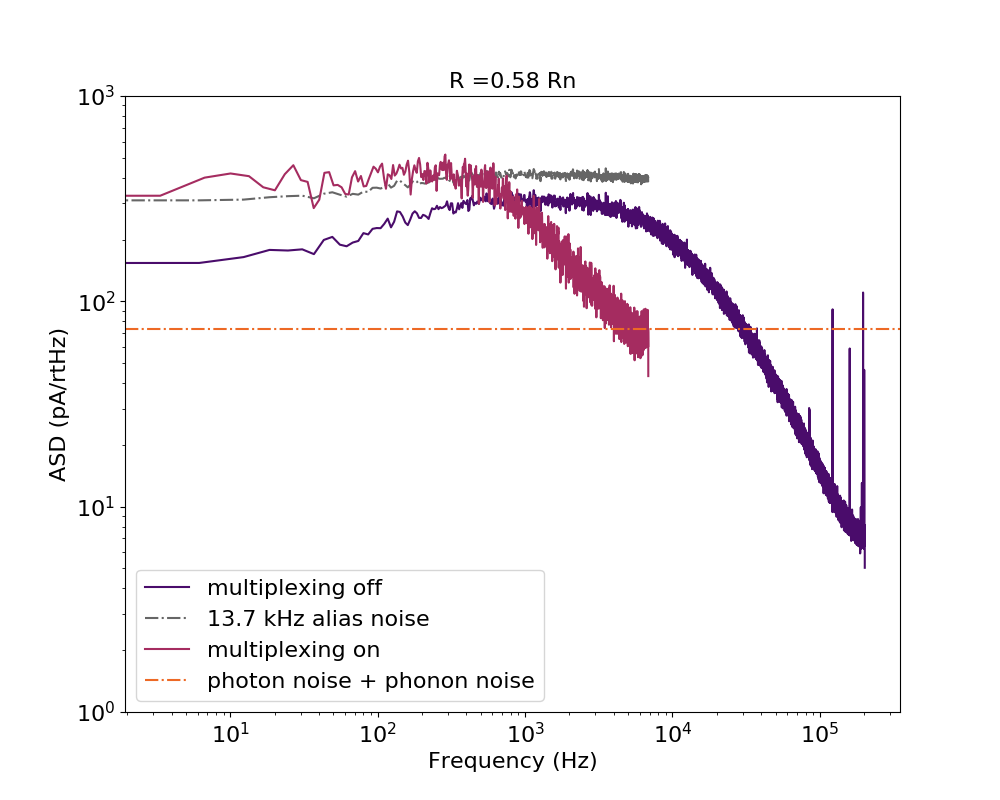}
    \caption{}
    \label{fig:alias_demo}
\end{subfigure}

\vspace{0.4cm}

\caption{
\textbf{(a)} Representative detector noise spectra measured at multiple operating points across the superconducting transition using high-sampling-rate ($\sim$400~kHz) readout with multiplexing disabled. A pronounced excess of high-frequency detector noise is observed, with the largest contribution occurring at lower in-transition bias points, where the TES transition is steepest.
\textbf{(b)} Single-detector noise spectra measured at $R/R_n = 0.58$. Noise spectra acquired with multiplexing enabled and disabled are compared. The aliased noise predicted from the multiplexing-off spectrum closely reproduces the multiplexing-on spectrum within the science band, demonstrating that the observed low-frequency excess noise arises from aliasing of high-frequency detector noise during TDM readout.
}
\label{fig:noise_alias}
\end{figure}

To characterize detector noise above the multiplexing bandwidth, high-sampling-rate ($\sim$400~kHz) noise timestreams were acquired with multiplexing disabled. Figure~\ref{fig:noise_psd} shows representative detector noise spectra measured at several operating points across the superconducting transition. A pronounced excess of high-frequency detector noise is observed, with the largest contribution occurring at lower in-transition bias points, where the TES transition is steepest.

To quantify the impact of this excess noise during normal operation, detector noise spectra acquired with multiplexing enabled and disabled were compared. The multiplexing-off spectra were used to predict the aliased detector noise expected during TDM readout. As shown in Figure~\ref{fig:alias_demo}, the predicted aliased spectrum closely reproduces the measured multiplexing-on spectrum within the science band, demonstrating that the elevated low-frequency detector noise is well explained by aliasing of excess high-frequency detector noise. 

\begin{figure}[H]
\centering
\includegraphics[width=0.7\textwidth]{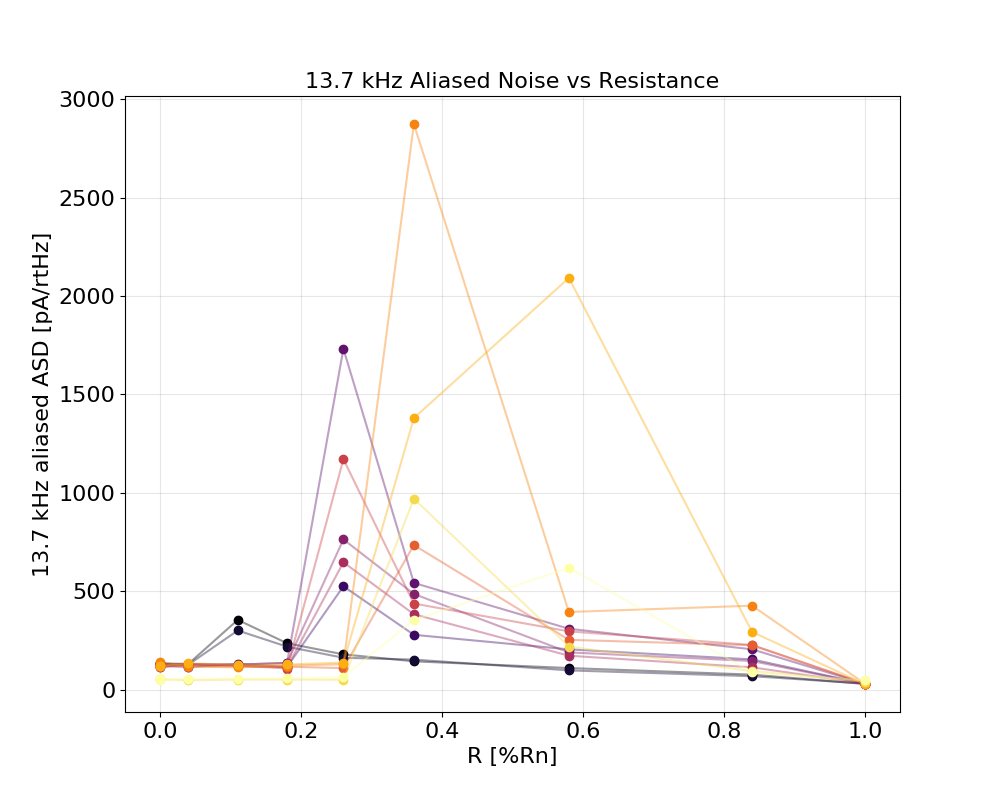}
\caption{Aliased noise contribution as a function of TES bias for representative detectors. Each curve corresponds to a different detector. The aliased noise peak occurs at different operating points within the superconducting transition, consistent with detector-to-detector variations in transition sharpness (\(\alpha\)) and the resulting electrothermal loop gain.}
\label{fig:alias_bias}
\end{figure}

The aliased noise contribution was further evaluated as a function of TES operating point for multiple detectors. Figure~\ref{fig:alias_bias} shows that the aliased noise reaches its maximum at different points within the superconducting transition for different detectors. This detector-to-detector variation suggests that the excess high-frequency noise is governed by intrinsic TES properties rather than the readout system alone, motivating direct measurements of the TES transition sensitivity.

This detector-to-detector variation suggests that the excess high-frequency noise is governed by intrinsic TES properties rather than the readout system alone. In particular, the dependence on operating point is consistent with variations in transition sharpness and electrothermal loop gain, motivating the direct measurements of the TES transition sensitivity presented in the following section.

\section{$\alpha$ Measurements}

The bias dependence of the aliased noise suggests that the observed excess noise is linked to intrinsic TES properties rather than the readout electronics. To investigate this possibility, direct measurements of the TES transition sensitivity, $\alpha$, were performed on representative BA2-150 and BA3-220/270 detectors fabricated using the inverted TES process \cite{Weber2024}.

The transition sensitivity was measured by gradually varying the focal-plane temperature while acquiring superconducting load curves. Detector resistance was extracted from the load curves and used to compute

\begin{equation}
\alpha = \frac{T}{R}\frac{dR}{dT}.
\end{equation}

Representative measurements of the TES transition sensitivity, $\alpha$, across the superconducting transition show that $\alpha$ values are significantly larger than the expected design values, implying elevated electrothermal loop gain\cite{Patel2026}.

The unusually large transition sensitivities are consistent with the sharper superconducting transitions introduced by the inverted TES fabrication process and provide independent experimental support for the interpretation that the excess high-frequency detector noise originates from detectors operating near their electrothermal stability boundary.


\section{Mitigation Strategies}

The physical interpretation presented in the previous sections suggests two operational approaches to reduce the aliased noise contribution: increasing the multiplexing rate and reducing the TES electrothermal loop gain.
Increasing the multiplexing rate raises the multiplexing Nyquist frequency, reducing the amount of excess high-frequency detector noise that aliases into the science band. In practice, however, the maximum achievable multiplexing rate is limited by the finite settling time of the SQUID readout electronics. To determine the fastest stable multiplexing configuration, SQUID settling-time measurements were performed and used to establish the minimum row length that could be reliably employed without introducing additional readout artifacts.
A second mitigation strategy exploits the dependence of loop gain on TES electrical power $P_e$. Raising the focal-plane temperature reduces the electrical power required to bias the TES within its superconducting transition, thereby lowering the electrothermal loop gain while maintaining the same operating resistance. Measurements performed at elevated bath temperatures show improved detector stability, reduced detector-to-detector scatter, and lower detector noise-equivalent temperatures (NETs), with improvements of approximately 10\% for both the BA2-150 and BA3-220/270 receivers.
Together, these results demonstrate that both optimized multiplexing parameters and reduced TES loop gain can substantially mitigate aliased detector noise in the current BICEP Array receivers while informing the design of future detector generations.

\begin{figure}[H]
\centering
\includegraphics[width=0.6\textwidth]{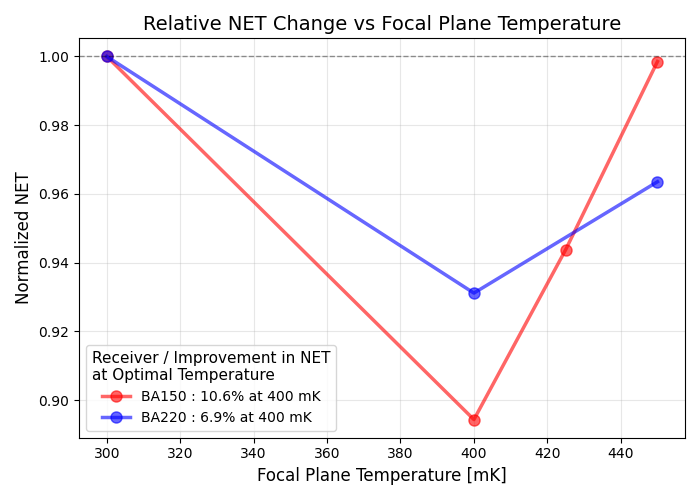}
\caption{Receiver normalized NET measured at the nominal operating resistance for several focal-plane temperatures. Increasing the focal-plane temperature reduces the receiver NET by lowering the TES electrical power and electrothermal loop gain. The receiver NET reaches a minimum near 400 mK and increases at higher temperatures as an increasing fraction of detectors saturate and can no longer contribute to the receiver sensitivity. The improvement is smaller for the BA3-220/270 receiver because the higher optical loading results in a larger photon-noise contribution, making the total receiver NET less sensitive to reductions in aliased detector noise.}
\label{fig:temperature}
\end{figure}

\section{Conclusions}

We have investigated the origin of the excess detector noise observed in the BICEP Array BA2-150 and BA3-220/270 receivers and demonstrated that the elevated detector noise-equivalent temperatures (NETs) are dominated by aliased high-frequency detector noise.
Direct measurements of the TES transition sensitivity, $\alpha$, reveal values significantly larger than expected, implying elevated electrothermal loop gains. Together with the observed dependence of the aliased noise on detector operating point, these measurements provide strong experimental support for the interpretation that the excess high-frequency noise originates from detectors operating near their electrothermal stability boundary.
Two operational mitigation strategies were demonstrated. Increasing the multiplexing rate reduces the amount of high-frequency noise aliased into the science band, while operating at elevated focal-plane temperatures lowers the TES electrical power and electrothermal loop gain. Together, these approaches improve detector stability and reduce the receiver NET by approximately 10\%.
These results establish elevated loop gain as the primary physical origin of the excess aliased noise in the BICEP Array high-frequency receivers and provide both operational and fabrication-level pathways for improving the performance of future TES detector arrays. Future detector designs will target reduced TES transition sharpness through transition engineering while preserving the magnetic robustness and fabrication advantages of the inverted TES architecture.

\section*{ACKNOWLEDGMENTS}
The BICEP/Keck experiments have been funded through U.S. National Science Foundation grants most re-
cently including 2220444-2220448, 2216223, 1836010, and 1726917. The research was carried out at the Jet
Propulsion Laboratory, California Institute of Technology, under a contract with the National Aeronautics and
Space Administration (80NM0018D0004). Focal plane development and testing were supported by the Gordon
and Betty Moore Foundation at the California Institute of Technology. Readout electronics were supported by
the Canada Foundation for Innovation grant to the University of British Columbia. The computations in this
paper were run on the Cannon cluster supported by the FAS Science Division Research Computing Group at
Harvard University. The analysis effort at Stanford University and the SLAC National Accelerator Laboratory
was partially supported by the Department of Energy. We thank the staff of the U.S. Antarctic Program and in
particular the South Pole Station without whose help this research would not have been possible. We also thank our winter-over operators: Manwei Chan, Karsten Look, Calvin Tsai, Paula Crock, Ta Lee Shue, Grantland Hall, Hans Boenish, Robert Schwarz, Sam Harrison, Anthony DeCicco, Thomas Leps, Brandon Amat, Nathan Precup, Steffen Richter, Thibault Romand, Danielle Simmons, Markus Ayasse, Steven Jungst, Nathan McReynolds, and John Della Costa.


\bibliographystyle{spiebib}
\bibliography{report_cap}

@article{BK2021,
  author = {{BICEP/Keck Collaboration}},
  title = {{Improved} {Constraints} {on} {Primordial} {Gravitational} {Waves} {using} {BICEP/Keck} {Observations} {through} {the} {2018} {Observing} {Season}},
  journal = {Physical Review Letters},
  volume = {127},
  number = {15},
  pages = {151301},
  year = {2021},
  doi = {10.1103/PhysRevLett.127.151301}
}

@incollection{IrwinHilton2005,
  author = {Irwin, Kent D. and Hilton, Gene C.},
  title = {{Transition-Edge} {Sensors}},
  journal = {Cryogenic Particle Detection},
  editor = {Enss, Christian},
  series = {Topics in Applied Physics},
  volume = {99},
  pages = {63--149},
  publisher = {Springer},
  address = {Berlin, Heidelberg},
  year = {2005},
  doi = {10.1007/10933596_3}
}

@article{Fatigoni2024,
  author = {Fatigoni, Sofia and others},
  title = {{Results} {and} {Limits} {of} {Time}-{Division} {Multiplexing} {for} {the} {BICEP} {Array} {High}-{Frequency} {Receivers}},
  journal = {Journal of Low Temperature Physics},
  year = {2024},
  note = {Proceedings of the 20th International Workshop on Low Temperature Detectors (LTD20)},
  doi = {10.1007/s10909-024-03174-0}
}

@article{Sonka2018,
  author = {Sonka, Rita and Bock, James J. and Megerian, Krikor G. and Steinbach, Bryan A. and Turner, Anthony D. and Zhang, Cheng},
  title = {{Characterization} {and} {Improvement} {of} {the} {Thermal} {Stability} {of} {TES} {Bolometers}},
  journal = {Journal of Low Temperature Physics},
  volume = {193},
  number = {5--6},
  pages = {877--884},
  year = {2018},
  doi = {10.1007/s10909-018-2009-z}
}

@article{Weber2024,
  author = {Weber, Alexis and others},
  title = {{Development} {of} {Inverted} {TES} {Bolometers} {for} {BICEP} {Array} {High-Frequency} {Receivers}},
  journal = {Journal of Low Temperature Physics},
  year = {2024},
  note = {Proceedings of the 20th International Workshop on Low Temperature Detectors (LTD20)}
}

@article{George2014,
  author  = {George, E. M. and Austermann, J. E. and Beall, J. A. and others},
  title   = {{A} {Study} {of} {Al-Mn} {Transition-Edge} {Sensor} {Engineering} {for} {Stability}},
  journal = {Journal of Low Temperature Physics},
  volume  = {176},
  number  = {3--4},
  pages   = {383--391},
  year    = {2014},
  doi     = {10.1007/s10909-013-0994-3}
}

@inproceedings{Hui2018,
  author = {Hui, Howard and Ade, P. A. R. and Ahmed, Z. and Aikin, R. W. and Alexander, K. D. and Barkats, D. and Benton, S. J. and Bischoff, C. A. and Bock, J. J. and Bowens-Rubin, R. and others},
  title = {{BICEP} {Array:} {A} {Multi-Frequency} {Degree-Scale} {CMB} {Polarimeter}},
  journal = {Millimeter, Submillimeter, and Far-Infrared Detectors and Instrumentation for Astronomy IX},
  editor = {Zmuidzinas, Jonas and Gao, Jian-Rong},
  series = {Proceedings of SPIE},
  volume = {10708},
  pages = {1070807},
  year = {2018},
  publisher = {SPIE},
  doi = {10.1117/12.2311725}
}

@misc{Patel2026,
  author       = {Anika Patel and others},
  title        = {{Engineering} {of} {Titanium} {Transition-Edge} {Sensor} {Wafers} {for} {the} {BA4} {90/150} {Receiver} {of} {BICEP} {Array}},
  howpublished = {Poster presented at SPIE Astronomical Telescopes + Instrumentation 2026},
  address      = {Copenhagen, Denmark},
  month        = jul,
  year         = {2026},
  note         = {}
}

\end{document}